\documentstyle[12pt]{article}

\begin{document}

\baselineskip=7mm

\newcommand{\be}{\begin{equation}}
\newcommand{\ee}{\end{equation}}
\newcommand{\bea}{\begin{eqnarray}}
\newcommand{\eea}{\end{eqnarray}}
\newcommand{\refs}[1]{(\ref{#1})}
\newcommand{\lag}{\langle}
\newcommand{\rag}{\rangle}
\newcommand{\ns}{\normalsize}

\begin{titlepage}
\title{{\Large\bf Gauged R-symmetry, Fermion and Higgs Mass Problem}\\
                          \vspace{-3.5cm}
                          \hfill{\ns IC/95/239\\}
                          \hfill{\ns hep-ph/9508320\\}
                          \vspace{3.5cm} }

\author{Eung Jin Chun \\[.5cm]
  {\ns\it International Center for Theoretical Physics}\\
  {\ns\it P.~O.~Box 586, 34100 Trieste, Italy}  }
\date{}
\maketitle
\begin{abstract}
\baselineskip=7mm  {\ns
We consider the simplest model of $SU(3) \times SU(2) \times U(1)_Y \times
U(1)_R$ gauge symmetry with one extra singlet field whose vacuum expectation
value breaks the horizontal $R$-symmetry $U(1)_R$ and gives rise to Yukawa
textures.
The $U(1)_R$ symmetry is able to provide both acceptable
fermion mass hierarchies and a natural solution to the $\mu$ problem
only if its mixed anomalies are cancelled by the Green-Schwarz mechanism.
When the canonical normalization $g_3^2=g_2^2={5\over3}g_1^2$ of the
gauge coupling constants is assumed, the Higgs mass parameter $\mu \sim
m_{3/2}$ can arise taking into acount the uncertainty in the ultraviolet
relation $m_e m_\mu m_\tau/m_d m_s m_b \simeq \lambda^q$ with $q \neq 0$.
When $q=0$ is taken only a suppressed value of $\mu \sim \lambda m_{3/2}$
can be obtained.
}\end{abstract}

\thispagestyle{empty}
\end{titlepage}

Recently gauged horizontal $U(1)$ symmetries  have been considered as an
appealing tool for understanding fermion mass problem \cite{ir}--\cite{dps}.
In the simpliest model with one extra $SU(3) \times SU(2) \times U(1)_Y$
singlet field, acceptable mass matrices arise if anomalies of a
horizontal $U(1)$ gauge symmetry is cancelled by the Green-Schwarz mechanism
\cite{gs}.  Remarkably the Green-Schwarz cancellation can predict the
ultraviolet value of the weak mixing angle $\sin^2\theta_W = 3/8$ \cite{iba}
without being related to grand unified groups.

Another persistent question in the minimal supersymmetric standard model
is the $\mu$-problem concerning the Higgs mass whose origin should be
related to physics beyond the standard model.
Even if supersymmetry can stabilize Higgs masses it alone does not explain
why the $\mu$-term $\mu H_1 H_2$ allowed by the standard model gauge symmetry
has such a small parameter $\mu$ compared to  e.g. the Planck mass: $\mu \ll
M_P$. It was pointed out \cite{nir,js2} that the horizontal $U(1)$ symmetry
yielding fermion mass hierarchies can resolve the $\mu$-problem by
forbidding the appearance of the direct $\mu$-term in superpotential and
allowing its effective generation through supersymmetry breaking
\footnote{The anomalous global Peccei-Quinn symmetry introduced
to solve the strong-CP problem can also be used to explain the appearance of
the $\mu$-term \cite{pq,cl}.}.
\bigskip

In this brief letter, we discuss how a horizontal $R$-symmetry $U(1)_R$
can provide  both fermion mass hierarchies and a natural generation of
the $\mu$-term.  It turns out that being compatible with fermion mass
matrices $U(1)_R$ has to be anomalous like non-$R$ horizontal symmetry.
Therefore, its spontaneous breaking gives rise to an unwanted axion
unless horizontal $R$-symmetries are gauged through the  Green-Schwarz
mechanism.  Gauged $R$-symmetry requires supersymmetry to be
local since $R$-symmetry does not commute with supersymmetry.

In the framework of local supersymmetry (supergravity) endowed with
$R$-invariance, the $\mu$-problem can be resolved in a natural
way \cite{kn}. If the Higgs mass term $H_1H_2$ carries $R$-charge zero,
it can appear only in K\"ahler potential:
\be \label{K} K \sim  H_1 H_2 \;, \ee
together with the usual kinetic terms of $H_{1,2}$.
This leads to the effective $\mu$-term in superpotential
\be \label{Weff} W_{eff} \sim {W \over M^2_P} H_1H_2 \;. \ee
In this way the appearance of the $\mu$-parameter is an inevitable
consequence of breakdown of both supersymmetry and $R$-symmetry,
through which the value of $\mu$ is determined by the gravitino mass
$m_{3/2} \equiv \lag W \rag/M_P^2$~: $\mu \sim m_{3/2}$.
Appearance of such a term in effective superpotential was discussed in general
supergravity theory \cite{mg} and in the context of superstring
theory \cite{cm}.  This situation has to be contrasted to the cases with
non-$R$ horizontal symmetry where the $\mu$-term has to appear through
non-renormalizable terms in K\"ahler potential.
In this case the $\mu$-parameter always has a suppression by factors
of the Cabibbo angle $\lambda \sim 0.2$: $\mu \sim \lambda^k m_{3/2}$ with
$k=1,2,\cdots$ \cite{nir,js2}.
\bigskip

Let us now discuss the conditions on horizontal $U(1)_R$ charges yielding
acceptable fermion mass matrices.  We assume the simplest case with only
{\it one} expansion parameter $\lambda$ resulting from the vacuum expectation
value of an extra singlet $\chi$: $\lambda \sim \lag \chi \rag/M_P$.
Let us denote the $R$-charges of $i$-th family quarks and leptons
in terms of the obvious notations: $q_i$, $u_i$, $d_i$, $l_i$ and $e_i$.
The $R$-charges of the corresponding squarks and sleptons are then
given by $q_i + 1$, etc.. We assign the $R$-charges $h_{1,2}$
for the Higgs fields $H_{1,2}$.  Therefore, the Higgsinos have the
$R$-charges $h_{1,2} -1$.  The  choice \refs{K} fixes $h_1 + h_2 = 0$.
The $R$-charge of the singlet is denoted by $-r_\chi$.
The superpotential of quarks and leptons invariant under $SU(3) \times SU(2)
\times  U(1)_Y \times U(1)_R$ can be written as
\be \label{Yuk} W \sim \lambda^{n^u_{ij}} Q_i U^c_j H_2
  + \lambda^{n^d_{ij}} Q_i D^c_j H_1 + \lambda^{n^e_{ij}} L_i E^c_j H_1
  \;. \ee
We follow the conventional normalization of $R$-charges: $R(W)=2$.
In eq.~\refs{Yuk}, the positive integers $n_{ij}^{u,d,e}$ are determined by
\bea \label{n's}
  q_i + u_i + h_2 - r_\chi n^u_{ij} &=& 0 \nonumber \\
  q_i + d_i + h_1 - r_\chi n^d_{ij} &=& 0 \\
  l_i + e_i + h_1 - r_\chi n^e_{ij} &=& 0 \nonumber \;.  \eea
When the charge assingments for the quarks and leptons do not allow
positive numbers $n_{ij}^{u,d,e}$, the corresponding Yukawa couplings are
absent.  This occurs whenever $(q_i + u_i + h_2)/r_\chi$ etc. are
{\it negative} or {\it non-integers}.
One can show that the determinants of the quark and lepton mass matrices
fulfill
\bea \label{det}
  \det M^u &\sim& v_2^3\lambda^{[\Sigma^u + 3h_2]/r_\chi} \nonumber\\
  \det M^d &\sim& v_1^3\lambda^{[\Sigma^d + 3h_1]/r_\chi} \\
  \det M^e &\sim& v_1^3\lambda^{[\Sigma^e + 3h_1]/r_\chi} \nonumber\eea
for {\it any values} of the $R$-charges \footnote{In
ref.~\cite{nir} it was noted that this relation holds for integer or
non-integer values of the $R$-charges.}.  Here $\Sigma^u \equiv
\sum_i(q_i+u_i)$, $\Sigma^d \equiv \sum_i(q_i+d_i)$, $\Sigma^e
\equiv \sum_i(l_i+e_i)$, and $v_{1,2} \equiv \lag H_{1,2} \rag$.
{}From eq.~\refs{det} we can draw two independent quantities
which can  be related to the anomalies of $U(1)_R$:
\bea \label{two}
  (\det M^u) (\det M^d) &\sim& v_1^3v_2^3 \lambda^{ [\Sigma^u + \Sigma^d +
        3(h_1+h_2)]/r_\chi } \nonumber\\
  (\det M^e)/(\det M^d) &\sim& \lambda^{ [\Sigma^e - \Sigma^d]/r_\chi }
   \;. \eea
The diagonalized fermion masses are known to satisfy the following
ultraviolet relations \cite{rrr}:
\bea \label{hierarchies}
  (m_u,m_c,m_t) &\sim& v_2(\lambda^8,\lambda^4,1) \nonumber\\
  (m_d,m_s,m_b) &\sim& v_1\lambda^x(\lambda^4, \lambda^2, 1) \\
  (m_e,m_\mu,m_\tau) &\sim& v_1\lambda^x (\lambda^{4}, \lambda^2, 1) \;.
  \nonumber\eea
We have put the same factor $\lambda^x$ for the $b-\tau$ unification:
$m_b \simeq m_\tau$. Note that $\tan\beta \equiv v_2/v_1 = \lambda^x m_t/m_b$.
Combining eqs.~\refs{two} and \refs{hierarchies} two restrictions on the
$R$-charges are obtained:
\bea \label{restrict}
  \Sigma^u + \Sigma^d &\simeq&  (3x+18+p) r_\chi - 3(h_1+h_2) \nonumber\\
  \Sigma^e - \Sigma^d &\simeq&  q r_\chi \;. \eea
Here two numbers $p,q$ take into account the uncertainties in the ultraviolet
mass relations \refs{hierarchies}: $|p|,|q| \leq 1$.
The two quantities in eq.~\refs{restrict} can be
expressed in terms of the anomalies $C_{3,2,1}$ of $U(1)_R$ with respect to
$SU(3)$, $SU(2)$ and $U(1)_Y$.
For the computation of the $U(1)_R$ anomalies we have to include the
contribution from gauginos (with $R$-charge 1)
as well as Hissinos, quarks and leptons.
The anomalies are given by \cite{cd}:
\bea \label{anomalies}
  C_3 &=& \sum_i(2q_i+u_i+d_i) + 6 \nonumber\\
  C_2 &=& \sum_i(3q_i+l_i) + (h_1-1) + (h_2-1) + 4 \\
  C_1 &=& \sum_i({1\over3}q_i+{8\over3}u_i+{2\over3}d_i+l_i+2e_i) +
                  (h_1-1) + (h_2-1) \;. \nonumber\eea
We do not consider the other anomalies like $U(1)_Y$--$U(1)_Y$--$U(1)_R$ or
$U(1)_R^3$  etc..  The former one entails specific $R$-charge assignments
for quarks and leptons and the latter requires full spectrum including hidden
supersymmetry breaking sector \cite{cd} which we do not address in this
paper. From eq.~\refs{anomalies} one gets two independent combinations
\bea \label{twoanomalies}
 C_3 &=& \Sigma^u + \Sigma^d + 6 \nonumber\\
 C_1 + C_2 - {8\over3}C_3 &=& 2\Sigma^e - 2\Sigma^d + 2(h_1+h_2) -16
  \;. \eea
Comparing eqs.~\refs{restrict} and \refs{twoanomalies}, one finds
 \bea \label{tworelations}
  C_3 &=& (3x+18+p) r_\chi - 3(h_1+h_2) + 6 \nonumber\\
  C_1 + C_2 - {8\over3}C_3 &=&  2qr_\chi + 2(h_1+h_2)-16 \;, \eea
from which the main results can be drawn.
\bigskip

In the case of $h_1+h_2=0$ corresponding to $\mu \sim m_{3/2}$ \refs{K}
two anomalies in eq.~\refs{tworelations} cannot vanish simultaneously within
the allowed uncertainties in $p$ and $q$.
Therefore one has to rely on the Green-Schwarz mechanism of anomaly
cancellation \cite{gs}.
The Green-Schwarz mechanism implies $\alpha_3 C_3 = \alpha_2 C_2 =
\alpha_1 C_1$ for the ultraviolet gauge coupling constants $\alpha_i =
g_i^2/4\pi$.  Thus the canonical normalization of the gauge couplings
$\alpha_3 = \alpha_2 = {5\over3}\alpha_1$ requires the second quantity in
eq.~\refs{tworelations} to vanish: $C_1+C_2-{8\over3}C_3=0$ \cite{iba}.
For this, one needs $q \neq 0$ and $r_\chi = 8/q$.

The $\mu$-term can result also from the following
non-renormalizable terms in the K\"ahler potential:
 \be \label{K'}
 K \sim {\chi^* \over M_P} H_1H_2 \quad\mbox{or}\quad
  {\chi\over M_P} H_1H_2 \;. \ee
The corresponding $\mu$-parameter is then given by $\mu \sim \lambda m_{3/2}$.
We will not consider the case with further suppressed values of $\mu$
since it becomes too small.  Two terms in eq.~\refs{K'} are allowed when
$h_1+h_2=k r_\chi$ with $k=\pm 1$, respectively.
Once again one finds that the anomaly cancellation can be achieved only by
the Green-Schwarz mechanism.  The canonical normalization requires the
$R$-charge $r_\chi = 8/(k+q)$.
Similar conclusions can be drawn also when $\lambda^2 \sim \lag \chi
\rag/M_P$ is taken as the expansion parameter.
\bigskip

In conclusion, we have shown that a gauged horizontal $U(1)$ $R$-symmetry
requires the Green-Schwarz mechanism for anomaly cancellation in order to
explain both the observed fermion mass hierarchies and the appearance of the
$\mu$-parameter.
The $\mu$-parameter comparable to the gravitino mass $m_{3/2}$ is
acceptable only if the ultraviolet mass relation $m_e m_\mu m_\tau/m_d m_s
m_b$ $\simeq$ $\lambda^q$ ($q=\pm1$) is assumed.
On the contrary, a suppressed value of $\mu \sim \lambda m_{3/2}$ is
consistent with any value of $q$.
Our conclusion is based on the minimal model with only one expansion parameter
generated by the vacuum expectation value of an extra singlet and on the
assumption of the canonical normalization of the gauge coupling unification
$g_3^2 = g_2^2 = {5\over3} g_1^2$.

\end{document}